\documentstyle[12pt]{article}  
\newcommand{\be}{\begin{equation}}
\newcommand{\ee}{\end{equation}}
\newcommand{\bea}{\begin{eqnarray}}
\newcommand{\eea}{\end{eqnarray}}
\newcommand{\bi}{\begin{itemize}}
\newcommand{\ei}{\end{itemize}}
\newcommand{\bc}{\begin{center}}
\newcommand{\ec}{\end{center}}
\newcommand{\bfl}{\begin{flushleft}}
\newcommand{\efl}{\end{flushleft}}
\newcommand{\bfr}{\begin{flushright}}
\newcommand{\efr}{\end{flushright}}

\def\6{\partial}

\def\={\!\!\!&=&\!\!\!}
\def\+{\!\!\!&&\!\!\!+~}
\def\-{\!\!\!&&\!\!\!-~}

\begin{document}
\title{Dualities in 4D Theories with Product Gauge Groups from Brane 
Configurations}
\author{{\em Radu Tatar,}\\
University of Miami, Department of Physics,\\
Coral Gables, Florida, USA\\
 tatar@phyvax.ir.miami.edu}
\maketitle
\begin{abstract}
We study brane configurations which correspond to N=1 field theories in four
dimensions. By inverting the order of the NS 5-branes
and D6-branes, a check on dualities in four dimensional theories can be made.
We consider a brane configuration which yields electric/magnetic duality
for gauge theories with $SO(N_{c1})\times Sp(N_{c2})$ product gauge group.
We also discuss the possible extension to any alternating product of SO and Sp 
groups. The new features arising from the intersection of the NS 5-branes
on the orientifold play a crucial role in our construction.
\end{abstract}
\newpage
\section{INTRODUCTION}
Three years ago, the study of non-perturbative effects in supersymmetric gauge
theories has received a big impact from the work of Seiberg and collaborators
\cite{sei1}\cite{sei2}.
They  studied the nonperturbative description of $N=1$ supersymmetric 
gauge theories for many gauge groups with flavors in different representations.

After the electric/magnetic dualities for $N=2$ theories were considered in 
\cite{sw}, Seiberg 
conjectured the electric-magnetic duality for $N=1$ gauge theories\cite{se1}.

The duality was formulated for gauge theories with gauge group $SU(N_{c})$ and
$N_{f}$ flavors in fundamental representation, the dual being a theory with
gauge group $SU(N_{f}-N_{c})$ and $N_{f}$ flavors in the fundamental 
representation. Following this example, an intense effort was made to obtain 
dualities for other gauge groups with flavors in all possible  representations.
\cite{kz1}\cite{kz2}\cite{sei3}\cite{sei4}\cite{sei5}\cite{ils}\cite{pou1}
\cite{pou2}

More recently, the connection between string dualities and field theory 
dualities
has become more and more evident. The most important tools are the D-branes,
which describe solitonic defects where the open string can end
\cite{pol2}. 

Because the open string can end on them,
they have a world-sheet Abelian gauge field. When we put N D-branes on top of 
each other, the system behaves like a U(N) gauge theory. By T-duality, we can 
navigate between lower and higher dimensional D-branes.
Besides D-branes, NS fivebranes are also important in obtaining the field 
theory dualities from string theories.

The first study of brane configuration came in the paper of Hanany and Witten
\cite{hw}.
They studied the brane configuration which gives N=2 in 4 dimensions.
Their construction is equivalent, by a T-duality transformation, to the
geometric singularity approach of \cite{sbv}\cite{ov1}\cite{ov}.
The main idea is to compactify type IIB  superstring on 
$K_{3}\times R^{5,1}$,
 $K_{3}$ being viewed as an elliptic fibration over the complex plane
and having singularities where one of the cycles in the fibre  shrinks to
zero. As shown in \cite{sbv}\cite{ov1},
by moving in the fibration between 2 singularities, the cycle covers a
holomorphic curve which has the property of breaking half of the supersymmetry.
So we can wrap a D-brane partially on this holomorphic curve and partially along
other noncompact direction in space time. In \cite{ov1}\cite{ov} it was proved 
that by 
taking the T-dual one obtains at  singularities
2 NS fivebranes carrying magnetic charge. The D p-brane which was wrapped around
the holomorphic curve in the original picture becomes now a D p-1 brane
with 1 compact direction, being stretched between 2 NS fivebranes.

For p=4, the result is a 3+1 worldvolume stuck between two NS fivebranes, 
with one compact direction, which is actually a 3 dimensional field theory.  
For $N_{c}$ D3-branes
on top of each other, one obtains a 3  dimensional theory with gauge group
$U(N_{c})$. If we insert $N_{f}$ D5-branes between the two NS fivebranes, 
  they describe a $U(N_{f})$ global theory because their worldvolume is
noncompact in all directions. By changing the positions of branes in spacetime,
very interesting connections have been obtained between Coulomb phase, Higgs
phase and the mirror symmetry.

In their very important paper, Elitzur, Giveon and Kutasov \cite{egk}
have used a configuration of branes which gives a theory with 4
supercharges in 4 dimensions. They have checked the Seiberg duality for the 
group $U(N_{c})$ with $N_{f}$ flavors in the fundamental representation.
The construction was generalized by other authors to other configurations.
Particularly interesting is the case of the SO and Sp groups, where the
string theory becomes non-orientable and we have to add orientifolds
\cite{pol2}\cite{eva}\cite{kuta}. More interesting results for brane configurations 
corresponding to three, four and five dimensional 
field theories have been obtained in 
\cite{ah}. The brane configurations were also discussed from the point of
view of strongly coupled string theory (M theory) and very nice connections with
Seiberg-Witten spectral curves have been obtained in \cite{wit}\cite{eva}. 
In the present work, we try to generalize the work of \cite{eva} and \cite{kuta}
to the case of product gauge group $SO(N_{c1})\times Sp(N_{c2})$, 
using results from the above papers and the result
of \cite{bsty} and \cite{bh} for the product gauge groups 
$SU(N_{c1})\times SU(N_{c2})$. The 
final
result will be compared with \cite{ils}.

In section two we will review some important aspects of the known dualities.

In  section three we will discuss a new duality given by brane configurations 
and we will see that the conjectures made in \cite{eva}\cite{kuta} are verified
in this case.

 We will conclude by making some comments.

\section{Some dualities from brane configurations}

\subsection{Brane configurations in the oriented case}
If one compactifies type IIA on a Calabi-Yau manifold, a configuration which 
preserves $N=1$ supersymmetry in 4 dimensions is obtained.
The Calabi-Yau manifold
is viewed as a double elliptic fibration. Take the cycles of the two tori 
living in the fiber in the (89) and (45) directions, wrap D-branes on
the holomorphic curves determined by moving these 2 cycles between two
singularity points and make the T dualities with
respect to the (45) directions and (89) directions to obtain two NS fivebranes
with different orientations.\footnote{This connection between the geometric 
singularity picture and the intersecting brane picture was very beautiful
explained by Prof. Ooguri in his excellent set of lectures held at Spring School
in String Theory-Trieste 1997.}

This is the configuration considered in \cite{egk} and consist in:

1)NS fivebrane with worldvolume $(x^{0}, x^{1}, x^{2}, x^{3}, x^{4}, x^{5})$,

2)NS' fivebrane with worldvolume $(x^{0}, x^{1}, x^{2}, x^{3}, x^{8}, x^{9})$,

3)D4 four branes with worldvolume $(x^{0}, x^{1}, x^{2}, x^{3}, x^{6})$,

4)D6 sixbranes with worldvolume $(x^{0},x^{1}, x^{2}, x^{3}, x^{7}, x^{8}, x^{9})$.

The 6-th direction is the compact direction for the D4 branes which is stretched
in that direction between the NS and NS' fivebranes. As explained in \cite{egk}
this brane configuration keeps 1/8 of the original SUSY, so it gives N=1 in four 
dimensions. The NS and NS'branes have to coincide in the 7-th direction, 
otherwise the supersymmetry will be further broken. 
In \cite{egk} one of the two NS branes was moved in the
7-th direction with the cost of breaking the $U(N_{c})$ gauge group given by
$N_{c}$ D4 branes living between NS and NS'. This is the Fayet-Iliopoulos
mechanism which protects us against supersymmetry breaking.

By moving the NS' brane to the left of the NS brane considered to be fixed,
the authors of \cite{egk} obtained the dual description. The main point was
the conservation of linking number (magnetic charge) which, as Hanany and Witten
first showed in \cite{hw} leads to the appearance of a new D4 brane everytime
NS' and a D6 branes change their positions. The dual description obtained after
all these interchanges describes a theory with gauge group $U(N_{f}-N_{c})$ and 
with the same number of flavors.

In the case of oriented string, the work of \cite{egk} was generalized in
\cite{bsty}
and \cite{bh} for the case of product groups $U(N_{c1})\times U(N_{c2})$.
\subsection{NONORIENTABLE CASE}
For nonorientable strings, the gauge group which appears when one
puts D-branes on top of each other is SO(N) or Sp(N). A new feature of
the nonorientable theories is the appearance of orientifolds. These are 
generalizations of the orbifolds, having a supplementary worldsheet symmetry
besides the space-time symmetry. Because of the orientifold, the D-branes are
forced to appear in pairs.
 The orientifold is a BPS state so it breaks half of the 
supersymmetry. To avoid a supplementary breaking of SUSY in the presence
of an orientifold, the authors of \cite{eva} and \cite{kuta} considered the case
of an orientifold which is parallel with the D branes. We can have 2 types of
orientifolds:

-O4 which are parallel and have the same worldvolume as the D4 branes.
 
-O6 which are parallel and have the same worldvolume as the D6 branes.

As a function of the orientifold charge, the configurations considered 
in \cite{eva}, \cite{kuta} can have SO or Sp gauge symmetries. By taking one
NS to be stuck on the orientifold and moving all other NS branes from the 
left to the right of the fixed one,  
a dual configuration is obtained and the result agrees
with the ones of field theory. The main point  is that during the
transition from electric to dual theory, the NS branes have to intersect  
and at the intersection point we are in strong coupling limit and some 
interesting  phenomena happen. The fact we are in the strong coupling
limit can be seen both from the point of view of string theory ( the dilaton
blows up there, so the string coupling constant becomes infinite) and
field theory ( the gauge coupling constant is proportional with the inverse
of the distance between the two NS branes we are talking about).

Another important aspect of the presence of the orientifold is that the
possible flavor symmetry is Sp when the gauge group is SO and SO when the
gauge group is Sp \cite{eva}. The origin of this difference 
is the different sign of
$\Omega^{2}$ when it acts on D4 and D6 branes, a fact that is
$T_{45789}$-dual to the situation of \cite{gp}\cite{gj} where the action 
of $\Omega^{2}$ had different signs when acting on D5 and D9 branes. 

\section{A NEW DUALITY}
We will consider in this paper the case of an O4 orientifold.

As we stated before, the orientifold is parallel with the D4 brane and the 
D4 brane is the
only brane which is not intersected by the O4 orientifold. 
The orientifold gives a spacetime reflection

$(x^{4},x^{5},x^{7},x^{8},x^{9})\rightarrow 
(-x^{4},-x^{5},-x^{7},-x^{8},-x^{9})$

which are all noncompact directions so the field line can go to infinity.

On directions where the orientifold is a point, any object which is extended
 along them will have a mirror copy of itself. The NS branes has a
mirror in $x^{4}, x^{5}$ directions, NS' has a mirror in $x^{8}, x^{9}$
and D6 has one in $x^{7}, x^{8}, x^{9}$ directions. These objects and 
their mirrors enter only once, taking both would be overcounting.
In our discussion, we will make a difference between the number of branes
when we count branes and their mirrors and the number of physical branes
when we count only branes without their mirrors. {\em We will specify
at any moment if we are referring to branes or physical branes}.

Another important aspect of the orientifold is its charge, given by the charge
of $H^{(6)}=d A^{(5)}$ coming from RR sector. In the natural normalization,
where the D4 brane carries unit charge, the charge of the O4 plane is 
$\pm 1$, for $-\Omega^{2}=\pm 1$ in the D4 brane sector.

We now introduce the electric theory. The gauge group is 
$SO(N_{c1})\times Sp(N_{c2})$
with $2N_{f1}$ flavors in the vector representation of SO group and $2N_{f2}$
flavors in the fundamental representation of Sp group. In brane language, this
corresponds to three NS branes . As discussed in
\cite{bh}, it is not sufficient to have only perpendicular branes like 
NS and NS' i.e.
in $(x^{4}, x^{5})$ and $(x^{8}, x^{9})$ directions. We need branes at different 
angles in $(x^{4}, x^{5}, x^{8}, x^{9})$ directions.  Denote them by A, B and C 
from right to left (so A is in the far right) on the compact $x^{6}$ direction.
Then we have the following orientation: the B brane is oriented at zero
degree, i.e in $(x^{4}, x^{5})$ direction, the brane A is oriented at
an angle $\theta_{1}$ with respect to B and C is rotated at an angle
$\theta_{2}$ with respect to B. {\em The angles $\theta_{1}$ and $\theta_{2}$
are not arbitrary.} The $N=1$ theory is obtained from $N=2$ supersymmetric theory
when we give mass to the adjoint fields of the $N=2$ theory. As explained in
\cite{bh}, the angles $\theta_{1}$ and $\theta_{2}$ are just given by:
\begin{equation}
m_{1} = \mbox{tan}(\theta_{1}), m_{2} = \mbox{tan} (\theta_{2})
\label{angle}
\end{equation}
where $m_{1}, m_{2}$ are the masses for the adjoint fields, one for SO groups and
the other one for Sp group, which are integrated out when we go from $N=2$ theory
to $N=1$ theory.  

From right to left we have: $N_{c1}$ D4 branes between A and B, $N_{c2}$
D4 branes between B and C . As a function of the orientifold projection,
we have two sectors, between A and B we have symmetric O4 projection
and between B and C we have antisymmetric O4 projection. 
Therefore the number of physical branes between A and B is $N_{c1}/2$ 
and the number physical branes between B and C
is $N_{c2}$. As discussed in \cite{eva}, the sign of 
the $A^{(5)}$ charge flips as one passes a NS fivebrane. If the sign of
$A^{(5)}$ is chosen to be positive between A and B, it will be negative between
B and C. For this reason the gauge group product $SO\times Sp$ or $Sp\times SO$
(if we choose the sign to be negative between A and B) is the only possibility,
meaning that $SO\times SO$ or $Sp \times Sp$ cannot exist. 
With our choice of sign, the gauge group is $SO(N_{c1})\times Sp(N_{c2})$.
Between A and B we have $2N_{f1}$ D6 branes which intersect the $N_{c1}$
 D4 branes
and between B and C we have $2N_{f2}$ D6 branes which intersect the $N_{c2}$ 
D4 branes. Here we count the number of branes, i.e. the number of branes plus
their mirrors. If we talk about the number of physical branes, we have
$N_{f1}$ physical D6 branes between A and B and $N_{f2}$ physical branes
between B and C. 

Strings stretching between the $N_{f1}$ physical D6-branes and the 
$N_{c1}/2$ physical D4 branes are the chiral multiplets 
in the vector representation
of $SO(N_{c1})$. Strings stretching between the $N_{f2}$ physical 
D6 branes and the
$N_{c2}$ physical D4 branes are the chiral multiplets in the fundamental 
representation of $Sp(N_{c2})$
group. The $N_{f1}$ physical D6 branes are parallel with the A brane and the 
$N_{f2}$ physical D6 branes are 
parallel with the C brane so there exist chiral multiplets which correspond 
to the motion of D4 branes in between the NS and D6 branes, 
as discussed in \cite{bh}. These states are precisely the chiral mesons of
the dual theory.
 
When one considers strings
stretched between the $N_{c1}/2$ and $N_{c2}$ physical 
D4 branes, a field X in the
($N_{c1},2N_{c2}$) representation of product gauge group is obtained. 

For the field X , there is a superpotential in the theory which truncates the
chiral ring. This superpotential is deduced as follows: start with an $N=2$
theory with gauge group $SO(N_{c1})\times Sp(N_{c2}), N_{f1}$ hypermultiplets
$Q^{i}$ 
 charged under $SO(N_{c1})$ and $N_{f2}$ hypermultiplets $Q'^{i}$ charged 
under $Sp(N_{c2})$
and the adjoint fields $X_{1}, X_{2}$ of respectively SO and Sp groups. 
Then we write the superpotential of the theory which has terms like
\begin{equation}
W = \lambda_{1} Q X_{1} Q + \lambda_{2} Q' X_{2} Q' + X X_{1} X + X X_{2} X
\end{equation}
Breaking the $N=2$ supersymmetry, we set $\lambda_{1} = \lambda_{2} = 0$,
we give the adjoint fields $X_{1}, X_{2}$ masses $m_{1}, m_{2}$ and
we integrate them out. What remains is a superpotential: 
\begin{equation}
W = -\frac{1}{2} (\frac{1}{m_{1}} + \frac{1}{m_{2}}) \mbox{Tr} X^{4}  
\end{equation}
So the superpotential of $N=1$ supersymmetric theory goes like
W=Tr$X^{4}$  and this truncates the chiral ring. As we discussed before
the masses of the adjoint fields are directly connected with the angle
of rotation on the NS fivebrane in $(x^{4}, x^{5}, x^{8}, x^{9})$ plane. 
as in equation (\ref{angle}).

Now we go to the magnetic theory.

\vspace{1cm}

{\em  We show that the result is a theory with the gauge group 
$SO(\tilde{N}_{c1})\times
Sp(\tilde{N}_{c2})$ with $\tilde{N}_{c1}=4N_{f2}+2N_{f1}-2N_{c2}$ and
$\tilde{N}_{c2}=2N_{f1}+N_{f2}-N_{c1}/2$. The anomaly cancellation for the
$Sp(N_{c2})$, requiring $N_{f2}+N_{c1}$ to be even, ensures that 
$\tilde{N}_{c2}$ is an integer.}

\vspace{1cm}

Like in \cite{bh}, first move all the physical 
$N_{f1}$  physical D6 branes (plus their mirrors if we talk about the total
number of branes) to the left of all NS branes. 
They are intersecting both B and C
NS branes. Using the linking number conservation argument,
it results that each physical D6 brane has two physical D4 branes on 
its right after transition.

By moving all $N_{f2}$ physical D6 branes to the right past, 
they intersect both
B and A and the conservation of linking number tells that each 
physical D6 brane has two physical D4 branes on its left.

Because the NS branes are trapped at the spacetime orbifold fixed points (which
form the orientifold plane), they cannot avoid the intersection 
so they have to 
meet and there is a strong coupling singularity. When each one of B and C 
actually meets A, such a singularity appears. From the field theory point of
view, such a non smooth behavior was expected because for Sp$(N_{c})$ and
SO$(N_{c})$ groups there is a phase transition.

In \cite{eva}, the effect of such a singularity was deduced
 to be the appearance
or disappearance of two D4 branes. In \cite{kuta} the transition 
was shown to be
smooth when the linking number in both sides of any NS brane is the same.

For SO groups, the procedure is to put two D4 branes on top of the orientifold
plane and break the other $N_{c1}-2$ D4 branes, entering in a Higgs phase. 
For Sp groups a pair of D4 branes and  anti-D4 brane plus their mirrors 
were created, the antifour-branes 
cancelling the charge difference along the orientifold.

Let us see how the transition to magnetic theory works. 
Remember that the initial configuration
is, from right to left: 

\vspace{0.5cm}

{\em the NS brane A, to its left the NS brane B ( between them the O4
projection being symmetric - SO), and to the left of B being the NS brane
C ( between B and C the O4 projection being antisymmetric - Sp).} 

\vspace{0.5cm}

First move C to the right of B. In \cite{eva} language, two D4 branes must
disappear because we have Sp group. Between B and C  branes we have a 
deficit of two D4 branes. Now move C to the right of A. 
When C passes A, 2 D4 branes appear between A and C because we have an 
SO group,
so we have now a deficit of 2 D4 branes between B and A and no deficit
between A and C. 
In this moment the configuration is as follows, from right to left:

the NS brane C, to its left the NS brane A (between them the O4 projection
being symmetric - SO) and to the left of A being the NS brane B ( between
A and B the O4 projection being antisymmetric - Sp and there is a deficit
of 2 D4 branes).

We want to move now B to the right of A in order to arrive to the
magnetic picture. We encounter a new phenomenon here. Between B and A
we have antisymmetric O4 projection, so after this transition,  
there is another deficit of 2 D4 branes between A and B ( from 
left to right). But the D4 branes which were before 
between B and A are changing the orientation after B comes to the right
of A so we have actually a deficit of 2 D4 branes with one orientation
and 2 D4 branes with another orientation which are thus cancelling each
other i.e. the addition of their physical charges gives 0. 
So there is no supplement or deficit of D4 branes. Remembering
that between A and C there was no supplement or deficit of D4 branes,
it results that there are no D4 branes which appear or disappear in the
transition from the electric to magnetic theory.

In  \cite{kuta} language, 
for a smooth transition
between B and C, we need to create 2 pairs of D4 branes and anti D4 branes, the
anti-fourbranes neutralize the charge difference along the orientifold. In the
same way as before, the 2 D4 branes to be put on top of the orientifold when
C passes A and B passes A annihilate the anti four branes. The two D4 branes 
which were on top of the orientifold came from the $N_{c1}$ fourbranes 
connecting A and B therefore leaving
$N_{c1}-2$ foubranes between A and B.
So the two D4 branes which remain after their anti-branes vanish
 add to the $N_{c1}-2$. Therefore, by smoothing the
transition we did not create any D4 or anti D4 branes, as expected from the 
field theory calculation.

The final picture is the following, from left to right:  

\vspace{1cm}

{\em $N_{f1}$ physical D6 branes 
connected by $2N_{f1}$ physical D4 branes with A. Between A and B we have 
$\tilde{N}_{c2}$ physical D4 branes, between B and C we have 
$\tilde N_{c1}/2$ physical
D4 branes and to the right of C we have $N_{f2}$ physical D6 branes, 
connected by $2N_{f2}$ physical D4 branes with C (plus their mirrors).}

\vspace{1cm}

Here, we have counted only the physical D branes. Between A and B we have 
antisymmetric O4-projection, so we are forced to place an even number of
D4 branes, the number of physical D4 branes being $\tilde{N}_{c2}$.
Between the $N_{f1}$ physical D6 branes and A we have $2N_{f1}$ physical
D4 branes because the $N_{f1}$ physical D6 branes have passed two NS branes
(B and C) so this is the correct number. The same for the $2N_{f2}$
physical D6 branes which are between the $N_{f2}$ physical D6 branes
and C.

We use the linking number of A to calculate $\tilde{N}_{c2}$ and we apply
the formula:
\begin{equation}
l_{NS} = \frac{1}{2} (R_{D6} - L_{D6}) + (L_{D4} - R_{D4})
+ Q(O4)(L_{O4} - R_{O4})
\end{equation}
where $(L, R)_{D6} ((L, R)_{O4}) [(L, R)_{D4}]$ is the number of 
physical D6 branes (O4 planes)
[physical D4 branes] to the left or right of the NS fivebrane for which we are 
calculating the linking number. Here Q(O4) is the charge of the
O4 plane. In the original picture, the A brane sees an O4 plane
of charge -1 on its left ( because of the symmetric O4 projection
between A and B ) and an O4 plane of charge +1 on its right. In the final
picture, the A brane sees an O4 plane of charge +1 on its right ( because of
the antisymmetric O4 projection ) and an O4 plane of charge -1 to its left.
So the contribution of the O4 plane to the linking number is the same
in the initial and in the final configurations. Therefore the conservation
of the linking number is the same as the conservation of the physical charge.  
In the original
picture the charge is $-N_{f1}/2-N_{f2}/2+N_{c1}/2$ where we used the
numbers of physical D branes. In the magnetic picture the charge is
$-N_{f1}/2+N_{f2}/2-\tilde{N}_{c2}+2N_{f1}$. We have counted only the
contributions of the physical branes. Making the above physical charges equal,
 we obtain
\begin{equation}
\label{eq1}
\tilde{N}_{c2}=2N_{f1}+N_{f2}-N_{c1}/2. 
\end{equation}
For the B brane, in the original and the final pictures it sees an O4 plane
of charge -1 to its left and an O4 plane of charge +1 to its right.
Therefore the conservation of the linking number is the same as the
conservation of the physical charge. This conservation between the initial
and the final configurations gives
(again we consider the number of physical branes):

\begin{equation}
-N_{f2}/2+N_{f1}/2+N_{c2}-N_{c1}/2=\tilde{N}_{c2}-\tilde{N}_{c1}/2-N_{f1}/2+
N_{f2}/2
\end{equation}

We obtain 
\begin{equation}
\label{eq2}
\tilde{N}_{c1}=4N_{f2}+2N_{f1}-2N_{c2}. 
\end{equation}

The values for
$\tilde{N}_{ci}, i=1,2$ coincide with the ones obtained in \cite{ils}. (for
$k=0$ case where $2k+1$ corresponds to the number of NS branes 
which are moving.
In this paper we considered the case of a single NS brane which is moving).

From the brane configuration discussed above, the field content of the theory
is:

\vspace{1cm}

{\em
-gauge group $SO(\tilde{N}_{c1})\times Sp(\tilde{N}_{c2})$,

-$N_{f2}$ fields in the vector representation of $SO(\tilde{N}_{c1})$,

-$N_{f1}$ fields in the fundamental representation of $Sp(\tilde{N}_{c2})$,

-a field Y in the ($\tilde{N}_{c1},2\tilde{N}_{c2}$) representation of the
product gauge group.

-the chiral mesons of the dual theory which appear as discussed before and 
which have the same form as in \cite{ils}.}

\vspace{1cm}

We will make an additional check of our result. Change the overall sign of 
$\Omega^{2}$ so the gauge group becomes now $Sp(N_{c1})\times SO(N_{c2})$.
Now we have $N_{c1}$ physical D4 branes between A and B because of the
antisymmetric O4 projection and
$N_{c2}/2$ physical D4 branes between B and C because of the symmetric
O4 projection. We have $2N_{f1}$ flavors in the fundamental
representation of the Sp group and $2N_{f2}$ in the vector representation
of the SO group. 

 When strings stretch 
between the $N_{c1}$ and $N_{c2}/2$ physical branes, the corresponding
field X is in the $(2 N_{c1}, N_{c2})$ representation and a
superpotential W=Tr$X^{4}$ appears. Going to the dual, the same manipulations
as above give us the following brane configuration (from left to right):

\vspace{1cm}

{\em
$N_{f1}$ physical D6 branes connected by $2N_{f1}$ 
physical D4 branes with A, $\tilde{N}_{c2}/2$ 
physical D4 branes connecting A and B, $\tilde{N}_{c1}$ physical D4 branes 
connecting B and C
and $N_{f2}$ physical D6 branes at the right of C which are connected by 
$2N_{f2}$ physical D4 branes with C.} 

\vspace{1cm}

Again we have $2N_{f1}$ physical D4 branes to the left of A and $2N_{f2}$
physical D4 branes to the right of C because the $N_{f1}$ physical D6 branes 
are passing
B and C and the $N_{f2}$ physical D6 branes are passing B and A.

During the transition, after C passes B and A, the intermediary configuration 
is similar with the one obtained in the
$SO(N_{c1})\times Sp(N_{c2})$ case, after moving C to the right of B and A.
This configuration is, from right to left: the NS brane C, the NS brane A
( with antisymmetric O4 projection between C and A) and the NS brane B
( with symmetric O4 projection between B and A and 2 supplementary
D4 branes between B and A). When we move B to the right of A, there are
2 supplementary D4 branes which appear between A and B but the 
previous supplementary
D4 branes are changing the orientation so they are cancelling each other.
Therefore there is no supplement or deficit of D4 branes.

We want to see which is the dual gauge group. Again the charges of the O4 plane
to the left and to the right of each NS brane is the same in the original
and in the dual theory so the conservation of the linking number is
the same as the conservation of the physical charge.
For A, the physical charge is $- N_{f1}/2 - N_{f2}/2 + N_{c1}$ 
in the original theory and
$-N_{f1}/2 + N_{f2}/2 - \tilde{N}_{c2}/2 + 2N_{f1}$ in the magnetic theory.
So we 
obtain $\tilde{N}_{c2} = 4N_{f1} + 2N_{f2} - 2N_{c1}$. The same condition for B
would give $\tilde{N}_{c1} = 2N_{f2} + N_{f1} - N_{c2}/2$.
From the field theory point of view, the initial theory can be viewed as 
$SO(N_{c2})\times Sp(N_{c1})$ with $2N_{f2}(2N_{f1})$ vector (fundamental)
flavors and the final theory can be viewed as 
$SO(\tilde{N}_{c2})\times Sp(\tilde{N}_{c1})$ with $2N_{f1}(2N_{f2})$ 
vector (fundamental) flavors. We see that the above obtained formulas
for the magnetic gauge group agree with (\ref{eq1}), (\ref{eq2}) for the 
choice of the electric gauge groups and flavor representations. 

This construction can be generalized to any product of gauge groups, 
but we have
to put them in alternating order i.e. $SO(N_{c1})\times Sp(N_{c2})\times 
SO(N_{c3})\times Sp(N_{c4})\times ...$. By changing the overall sign of
$\Omega^{2}$ we can start with a Sp group from right to left.

For a product of more than 2 gauge groups, there are two cases:

-when there is an even number of gauge groups in the product, the effects of SO
and Sp projections will cancel each other so in the overall
picture of the dual no D4 branes appear or disappear. The result is similar to 
the
one obtained by \cite{bh} with the modifications that are to be done when
one considers non-orientable string theory. Taking their results, we modify
$\tilde{N}_{c}$ to $2\tilde{N}_{c}$ and $N_{c}$ to $2N_{c}$ anytime we talk
about the Sp gauge groups, obtaining the dual for the alternating product
of SO and Sp gauge groups. The argument that we use for the product of
2 gauge groups applies also here. So one has to be careful when changing
the positions of 2 NS branes connected by supplementary D4 branes or having
a deficit of D4 branes between them. 

-when there is an odd number of gauge groups in the product, we need to 
create or to annihilate D branes in the overall picture 
(or to put D4 branes or anti D4 branes on
the top of the orientifold in order to make a smooth transition).

The final result should be checked by field theory methods.
\section{CONCLUSIONS}
In this paper, by changing brane positions in space time, we managed to check
the duality for a theory with $SO\times Sp$ gauge groups and matter
in vector and fundamental representation. We  also discussed the generalization
to the case of products of more than 2 gauge groups.

A very nice check of the results of brane configuration picture
would be to obtain the same results in the geometric singularity picture,
where dualities have been checked for simple SU, SO and Sp groups, but not for 
gauge group 
products \cite{ov1}\cite{corea}. Also, a lot of results obtained in field
theory dualities remain to be rederived and verified from brane configurations.

\section{ACKNOWLEDGEMENTS}
We would like to thank Hiroshi Ooguri for his very interesting set of lectures
at Trieste-97, which helped to clarify our understanding. We would like
to thank David Kutasov and  Clifford V. Johnson for very important 
comments on the 
manuscript and for many explanations and Orlando Alvarez for advices. We
 would like to express our gratitude 
to the organizers of Spring School 1997 (Trieste) for their intense effort
to organize an extremely interesting school.

\end{document}